\documentclass[runningheads]{llncs}

\usepackage[T1]{fontenc}
\usepackage[english]{babel}

\usepackage{amsthm}

\usepackage{lmodern}
\usepackage{amsfonts}
\usepackage{mathtools}
\usepackage{verbatim}
\usepackage{listings}
\usepackage{dsfont}

\usepackage{pstricks}
\usepackage{pstricks-add}
\usepackage{pst-plot}
\usepackage{pst-text}

\usepackage{tikz}
\usepackage{tkz-tab}
\usetikzlibrary{arrows}

\usepackage[framemethod=tikz]{mdframed}
\usepackage{stmaryrd}
\usepackage{multicol}
\usepackage{sectsty}
\usepackage{enumerate}
\usepackage{ytableau}
\ytableausetup{centertableaux}
\usepackage{tcolorbox}
\usepackage{systeme}

\usepackage{graphicx}
\usepackage{hyperref}

\theoremstyle{plain}
\newtheorem{THM}{Theorem}

\theoremstyle{definition}

\newtheorem{exemple}[THM]{Example}

\theoremstyle{remark}

\begin{document}

\title{Generating Random Hyperfractal Cities}

\author{Geoffrey Deperle\inst{1} \and
Philippe Jacquet\inst{1}}

\authorrunning{G. Deperle and P. Jacquet}

\institute{INRIA Paris-Saclay, Palaiseau, France\\
\email{geoffrey.deperle@inria.fr}}

\maketitle

\begin{abstract}
This paper focuses on the challenge of interactively modeling street networks.
In this work, we extend the simple fractal model, which is particularly useful
for describing small cities or individual districts, by constructing random cities
based on a tiling structure over which hyperfractals are distributed. This
approach enables the connection of multiple hyperfractal districts, providing a
more comprehensive urban representation. Furthermore, we demonstrate how this
decomposition can be used to segment a city into distinct districts through
fractal analysis. Finally, we present tools for the numerical generation of
random cities following this model.

\end{abstract}

\section{Introduction}

In order to provide usable datasets for artificial intelligence applications, it
is often necessary to generate random cities with adjustable parameters, such as
the spatial extent of the city and its traffic distribution.

The first model describing the geometry of a city network is the fractal model
\cite{batty1994fractal}. Indeed, a city can be represented by a network that is
self-similar in the sense that big avenues split into large streets, which in
turn split into smaller paths. The model is incomplete as it describes only the
geometry of the roads but not their traffic. The hyperfractal model, as
introduced by Jacquet et al.\ in \cite{jacquet:hal-01498987}, provides a unified
framework where both the geometry and the distribution of traffic can be
described in a self-similar and hierarchical manner
\cite{popescu2021characterizingenergytradeoffsendtoend}. It offers a powerful
way to simulate cities where traffic is concentrated on main roads and
progressively diluted along increasingly finer streets. This distribution is
well-known as a power-law distribution:
\[
\text{frequency} \propto \frac{1}{(\text{rank})^{a}}.
\]

In this work, we extend the hyperfractal framework to model full cities composed
of interacting neighborhoods, each described by its own hyperfractal geometry.
By combining tiling techniques with recursive traffic models, we introduce a new
class of synthetic cities that better mimic the complexity of real urban
systems.

\section{Hyperfractal Model}

\subsection{Hyperfractal Model: The General Case}

\begin{definition}\label{def:local-dimension}
Let $\mu$ be a Borel measure on a compact metric space $(X,d)$ where $d$ is a
distance (usually the Euclidean distance), and let $x \in X$. Suppose there
exists $D > 0$ such that
\[
\lim_{\rho \to 0} \frac{\ln \mu(B_\rho(x))}{\ln \rho} = D.
\]
Then $D$ is called the \emph{local dimension} of $\mu$ at $x$.
\end{definition}

\begin{exemple}
Let $X = [0,1]^D$ be the $D$-dimensional unit hypercube, and let $\mu$ be the
Lebesgue measure on $[0,1]^D$. Then, for every point $x \in [0,1]^D$, the local
dimension of $\mu$ at $x$ is equal to $D$, since the volume of a Euclidean ball
$B_\rho(x)$ is proportional to $\rho^D$.
\end{exemple}

\begin{definition}\label{def:dim-measure}
Let $\mu$ be a measure supported on a subset $A \subset X$ (which we will later
interpret as our network). If the local dimension of $\mu$ exists and is equal
to $D$ at every accumulation point $x \in A \setminus \partial A$, then we say
that $\mu$ has (global) dimension $D$.
\end{definition}

\begin{definition}\label{def:hyperfractal}
The measure $\mu$ is said to be \emph{hyperfractal} if $D > 2$, i.e., if the
dimension of the measure exceeds the dimension of its support.
\end{definition}

Although it may seem paradoxical, a measure can have a dimension larger than
that of its support when its mass becomes extremely concentrated in a highly
irregular way at small scales. This ``explosive'' behavior, often seen in
recursive or random constructions, causes the measure to locally mimic a
higher-dimensional distribution despite being supported on lower-dimensional
structures.

It is well-known that in the setting of iterated function systems (IFS), one
can compute the dimension of a self-similar measure in an explicit way. These
measures are natural candidates when modeling recursive mass distributions,
including in fractal and hyperfractal settings.

\begin{THM}[Dimension of a self-similar measure {\cite{geronimo}}]\label{thm:geronimo}
Given a finite family of contraction similarities
$f_i(x) = \lambda_i x + t_i$, $i \in I$, with $|\lambda_i| < 1$, and a
corresponding probability vector $(p_i)_{i \in I}$, there is a unique Borel
probability measure $\mu$ such that
\[
\mu = \sum_{i \in I} p_i\, f_i \mu.
\]
The dimension of $\mu$ is then given by
\[
\dim(\mu)
= \frac{\displaystyle\sum_{i \in I} p_i \log \left(\frac{1}{p_i}\right)}
       {\displaystyle\sum_{i \in I} p_i \log \left(\frac{1}{\lambda_i}\right)}
= \frac{\text{entropy}}{\text{Lyapunov exponent}}.
\]
\end{THM}

Computing the hyperfractal dimension is essential for solving certain urban
network problems, such as the performance analysis of ad hoc networks, where
modeling the city using its hyperfractal dimension as input data is crucial
(e.g.\ \cite{popescu2019informationdisseminationspeeddelay}).

For clarity and to capture the essential features of many natural and urban
networks, we restrict our attention to a self-similar setting where all segments
are scaled by a fixed length ratio and the measure follows a uniform
distribution with geometric scaling. This simplified model already exhibits rich
behavior and serves as a foundation for more general constructions.

\begin{definition}\label{def:self-similar-network}
Let $(X,d) \subset \mathbb{R}^n$ be a metric space consisting of a countable
collection of line segments constructed recursively. This network is said to be
\emph{self-similar} if, at each step $n \geq 0$, a finite collection of segments
$\{S_{n,i}\}_{1 \le i \le N_n}$ is added, each of length
\[
\ell_n = c \, s^n
\]
for some scaling factor $0 < s < 1$ and constant $c > 0$. The network
$\mathcal{A}$ is then defined as the union of all segments at every step:
\[
\mathcal{A} = \bigcup_{n \ge 0} \bigcup_{i = 1}^{N_n} S_{n,i}.
\]
\end{definition}

\begin{definition}\label{def:uniform-self-similar-measure}
A \emph{uniform self-similar measure} on the network $\mathcal{A}$ is a Borel
measure $\mu$ satisfying the following:
\begin{enumerate}
  \item $\mu$ is supported on the union of the segments $\{S_{n,i}\}$;
  \item on each segment $S_{n,i}$, the measure is uniformly distributed:
  \[
    \mu_{n,i} = \frac{m_n}{\ell_n} \,\lambda_{S_{n,i}},
  \]
  where $\lambda_{S_{n,i}}$ denotes the Lebesgue measure restricted to the
  segment;
  \item the mass $m_n$ assigned to the segments at step $n$ follows a geometric
  growth or decay:
  \[
    m_n = m_0\, r^n, \qquad r > 0.
  \]
\end{enumerate}
\end{definition}

\begin{THM}\label{thm:dimension-uniform-ss}
Let $\mu$ be a uniform self-similar measure on a network as defined above, with
$\ell_n = c\, s^n$ and $m_n = m_0\, r^n$. Then, for $\mu$-almost every point
$x$, we have
\[
\dim(\mu)(x) = \log_s r = \frac{\ln r}{\ln s}.
\]
\end{THM}


\newpage

\subsection{Hyperfractal Manhattan's Model}

We introduce a toy model: the \emph{Hyperfractal Manhattan model}, which
captures the hyperfractal geometry of traffic in New York City's Manhattan
borough. It defines a measure $\mu$ supported on an infinitely resolved street
grid. At level $0$, thick lines form the initial grid. Each subsequent level
recursively subdivides the map into four regions, adding finer lines with scaled
density, creating an increasingly detailed and hierarchical structure.

\begin{figure}[!h]
  \centering
  \includegraphics[width=2.5cm]{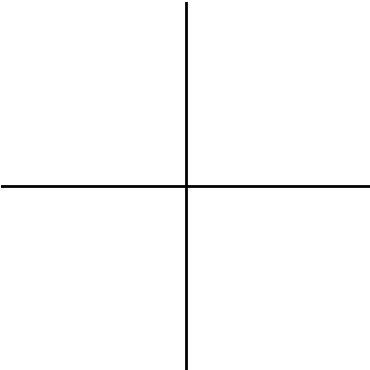}
  \includegraphics[width=2.5cm]{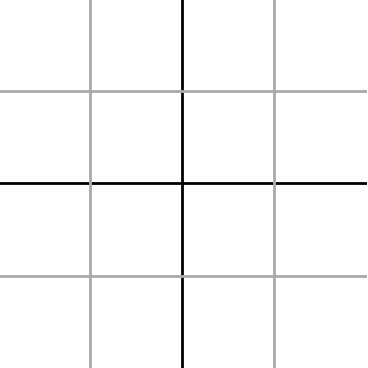}
  \includegraphics[width=2.5cm]{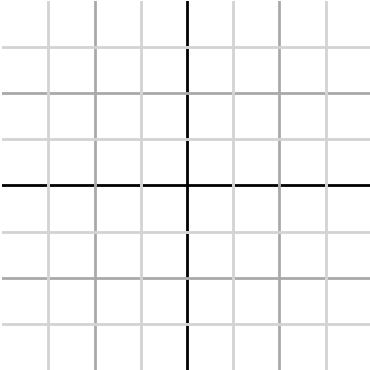}
  \includegraphics[width=2.5cm]{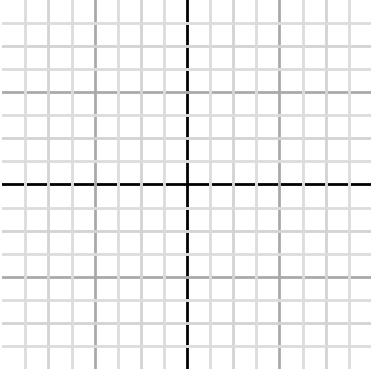}
  \caption{Recursive procedure for constructing the Manhattan model.}
  \label{fig:manhattan-grid}
\end{figure}

Let us denote this structure by
\[
X = \bigcup_{l = 0}^{+\infty} X_l
\]
with
\[
X_l = 
\bigl\{(b 2^{-(l+1)},y) : b = 1,3,\dots,2^{l+1}-1,\; y \in [0,1]\bigr\}
\cup
\bigl\{(x,b 2^{-(l+1)}) : b = 1,3,\dots,2^{l+1}-1,\; x \in [0,1]\bigr\},
\]
where $l$ denotes the level and $b$ is an odd integer (see Fig.~\ref{fig:manhattan-grid}).

We denote this structure by $\mathcal{A}$. Each segment has a depth $k$ in the
structure. There are $2 \times 4^k$ segments of depth $k$. Each segment of
depth $k$ has length $2^{-k}$ and density $\dfrac{p}{2}\left(\dfrac{q}{4}\right)^k$.
One can verify that $\mu$ is a probability measure:
\[
\mu(\mathcal{A})
= \sum_{e \in \mathcal{A}} \mu(e)
= \sum_{k=0}^{+\infty} 2 \times 4^k \times \frac{p}{2}\left(\frac{q}{4}\right)^k
= p \sum_{k=0}^{+\infty} q^k = 1.
\]

\begin{figure}[!t]
  \centering
  \includegraphics[width=3.5cm]{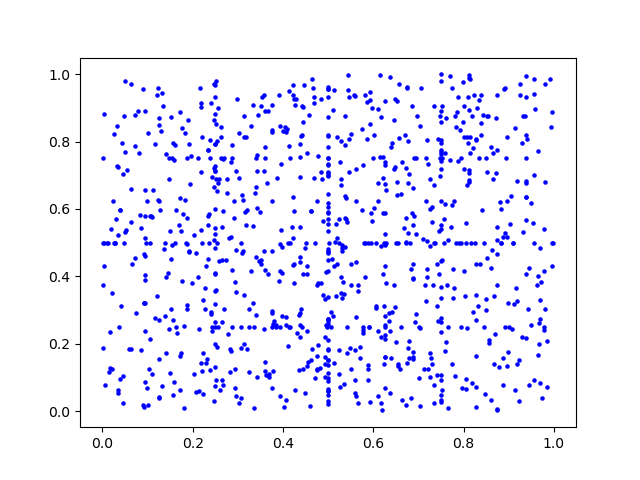}
  \includegraphics[width=3.5cm]{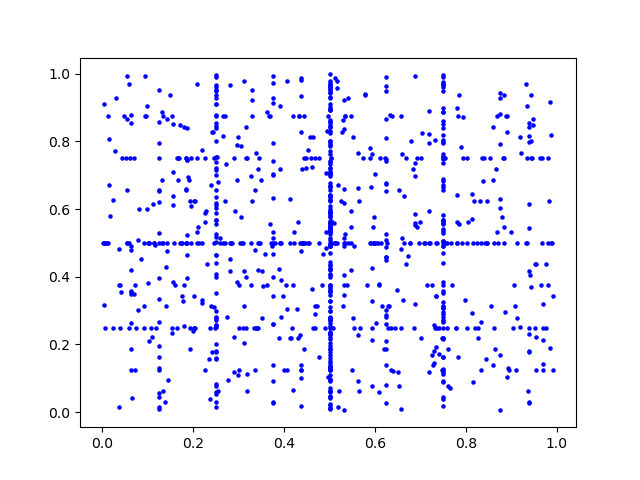} \\
  \includegraphics[width=3.5cm]{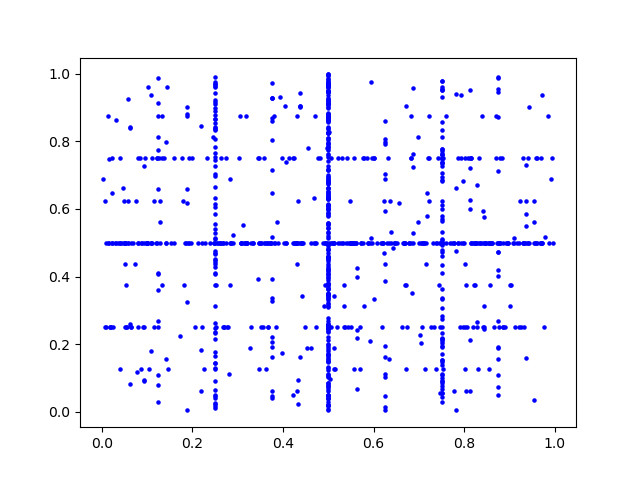}
  \includegraphics[width=3.5cm]{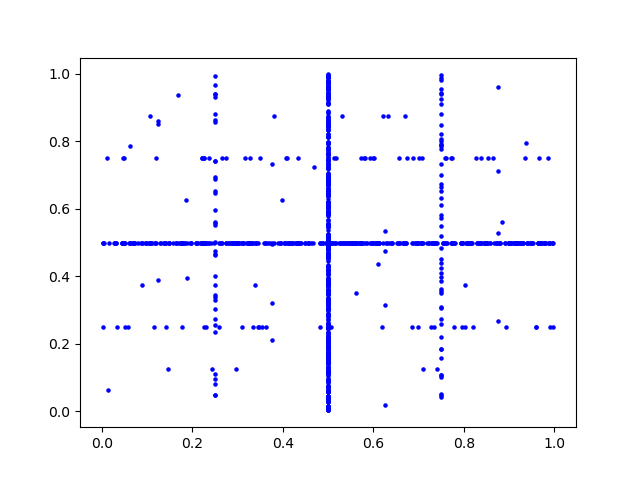}
  \caption{Example of $1000$ points generated randomly on the Manhattan grid
  with the measure $\mu$ for the parameters $p = 0.1, 0.3, 0.5, 0.8$.}
  \label{fig:manhattan-samples}
\end{figure}

The measure of the Manhattan model is self-similar and, by applying
Theorem~\ref{thm:dimension-uniform-ss}, we find that
\[
\dim(\mu) = \frac{\ln\left(\dfrac{4}{q}\right)}{\ln 2}.
\]
One can verify that for every $p \in [0,1]$, $\dim(\mu) > 2$. The measure is
therefore hyperfractal. As $p \to 0$, the measure $\mu$ approaches a uniform
distribution over $[0,1]^2$, indicating evenly spread traffic and a fractal
dimension $\dim(\mu)$ close to $2$. Conversely, as $p \to 1$,
$\dim(\mu) \to +\infty$, meaning traffic concentrates on a few dominant
streets, reflecting a highly hierarchical network structure (see
Fig.~\ref{fig:manhattan-samples}).

\subsection{Estimation of the Hyperfractal Dimension}

Suppose we have a hyperfractal measure $\mu$. How can we compute the dimension
of $\mu$ from observations of length and weights of a finite number of edges?

One can use the following theorem (see
\cite{popescu2019informationdisseminationspeeddelay}).

\begin{THM}\label{thm:nu-xi}
Let $S$ be a segment and $\mu(S)$ be the density of the segment, and let
$C_\ell(S)$ denote the accumulated length of the segments that have a larger
density than $\mu(S)$. For $\xi > 0$, let
\[
\nu(\xi) = \mu\bigl(C_\ell^{-1}(\xi)\bigr),
\]
where $C_\ell^{-1}(\xi)$ is the road segment $S$ with the smallest density such
that $C_\ell(S) \le \xi$. The hyperfractal dimension $\dim(\mu)$ appears in the
asymptotic estimate of $\nu(\xi)$ when $\xi \to +\infty$ with
\[
\nu(\xi) = O\bigl(\xi^{1 - \dim(\mu)}\bigr).
\]
\end{THM}

For the Manhattan model, if $S_k$ is a segment of order $k$, the cumulative
distance $C_\ell(S_k)$ is given by $2^{k+1} - 1$. Thus $k$ is of order
$\dfrac{\ln(\xi)}{\ln 2}$ and
\[
\nu(\xi) = O\left(\xi^{\frac{\ln\left(\frac{q}{2}\right)}{\ln 2}}\right),
\]
which can be rewritten as $O\bigl(\xi^{1 - \dim(\mu)}\bigr)$ since
$\dim(\mu) = \dfrac{\ln(4/q)}{\ln 2}$.

The procedure for the computation of the fractal dimension has the following
four steps:
\begin{enumerate}
  \item Collect street length and average annual traffic data.
  \item Subdivide each street into consecutive segments, ensuring the density
  variation between them is bounded by a fixed factor $A > 1$. In the ideal
  case ($A = 1$), all segments have equal density.
  \item Rank the streets by decreasing density
  $\lambda_1 \ge \lambda_2 \ge \dots$, and compute the cumulative length
  accordingly.
  \item Plot the density as a function of cumulative length and fit a power
  law to estimate the fractal dimension $\dim(\mu)$.
\end{enumerate}

\section{Tessellation and Hyperfractal Model}

The Manhattan model is relevant for representing the city of Manhattan and
cities where there is just one district. However, a city can be represented not
by one, but by a union of neighborhoods that are fractals following a
hyperfractal model.

\subsection{The Model}

Let us consider the unit square $[0,1]^2$ and a finite convex tiling of the
square. Each element of the tiling is called a cell. Let
$C_1,\dots,C_n$ be the different cells of the unit square. The idea is to draw
a hyperfractal network on each cell.

\begin{definition}\label{def:fractal-city}
We define a \emph{fractal city} as a finite union of networks
$\mathcal{A}^1,\dots,\mathcal{A}^n$ that we call neighborhoods, with measures
$\mu_1,\dots,\mu_n$ that are hyperfractal measures such that, for each
$i \in \{1,\dots,n\}$, $\mu_i$ is a hyperfractal measure on
$\mathcal{A}^i$.
\end{definition}

\begin{figure}[!h]
  \centering
  \includegraphics[width=6cm]{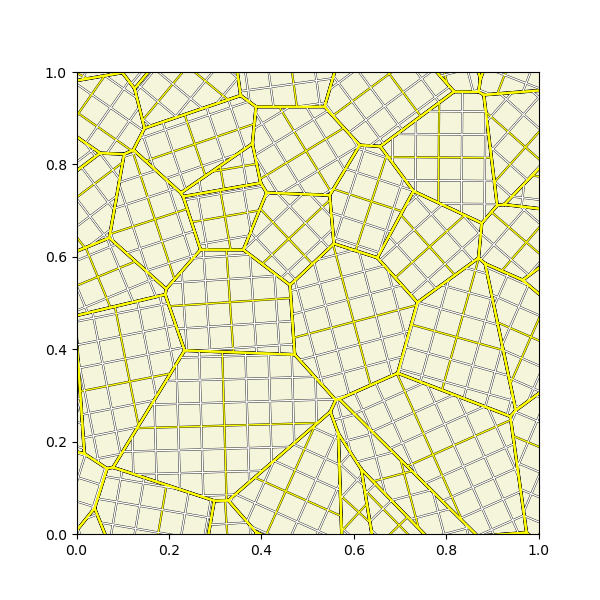}
  \caption{Example of a fractal city with a Voronoi tiling where each network
  is a Manhattan grid.}
  \label{fig:city-voronoi}
\end{figure}

To avoid overlaps of networks, we assume that each network is embedded in a
convex set so that the fractal city is composed of a convex tiling where, in
each cell $C_1,\dots,C_n$, there corresponds a network
$\mathcal{A}^1,\dots,\mathcal{A}^n$ and a measure
$\mu_1,\dots,\mu_n$ (see Fig.~\ref{fig:city-voronoi}).

In order to consider a measure on the fractal city as a whole, let us consider
$p_0 \in [0,1]$ as the weight of the roads that split the city into neighborhoods
(the edges of the diagram). To each cell $C_i$ we associate a weight $q_i$ such
that
$
\sum_{i=1}^n q_i = 1 - p_0$
which will be the total weight of cell $C_i$.

On each cell, we assign hyperfractal measures $\mu_1,\dots,\mu_n$ on
$\mathcal{A}^1,\dots,\mathcal{A}^n$ such that the total measure of each
neighborhood is $q_i$.

\section{Generating Random Fractal Cities}

We observe that in large North American cities, the main roads (often highways)
delineate neighborhoods which can themselves be modeled using a hyperfractal
model such as the Manhattan model.

Let us consider a model with $n$ neighborhoods represented by their centers
$x_1,\dots,x_n \in [0,1]^2$. Let $p_0$ be the total density of the diagram
edges.

Each neighborhood is modeled by a Manhattan model where the main axes have a
total density $\lambda_i p_i$ and each quadrant is a replica with a scale factor
$\dfrac{1 - p_i}{4}$.

\subsection{Generating Cities with This Model}

Given $n$ neighborhoods with centers $x_1,\dots,x_n$, weights
$\lambda_1,\dots,\lambda_n$, and scale factors $p_1,\dots,p_n$, a city can be
generated by placing these centers and assigning a hyperfractal model to each
Voronoi cell. Instead of a uniform layout, neighborhood centers tend to cluster
near the city center. This behavior can be modeled using a Gaussian
distribution with a covariance matrix encoding the degree and direction of
urban sprawl.

\subsection{Urban Sprawling}

Urban expansion is often constrained by geographical features, leading to
anisotropic growth patterns. In particular, cities may develop preferentially
along natural elements such as rivers, coastlines, or mountain ranges. To model
this phenomenon, the distribution of neighborhood centers can be represented
using a Gaussian distribution, with its variance (covariance) matrix structured
to reflect the directional characteristics of expansion.

\paragraph{Gaussian Distribution for Neighborhood Centers.}

\begin{figure}[!h]
  \centering
  \includegraphics[width=4cm]{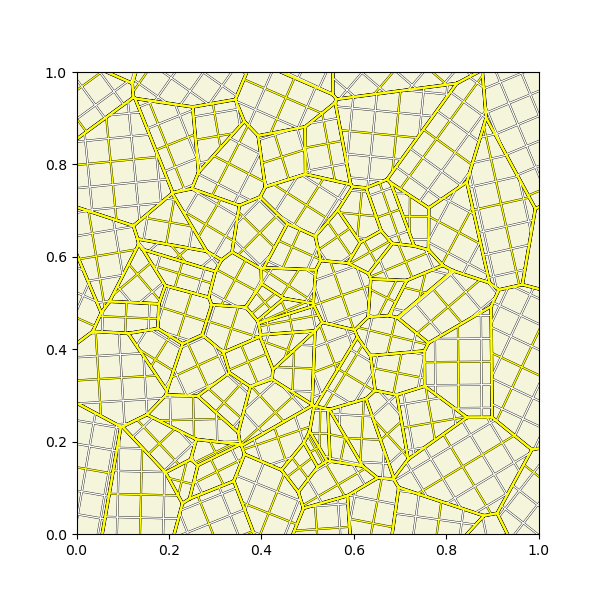}
  \includegraphics[width=4cm]{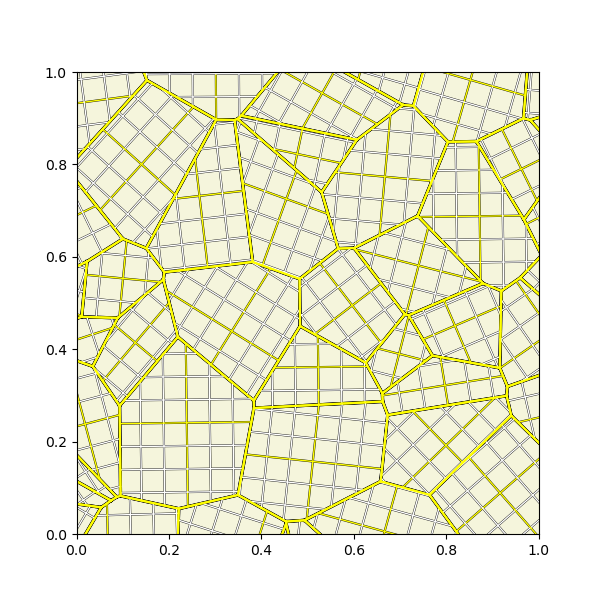}
  \caption{Example of cities generated with covariance matrices
  $\begin{pmatrix} 0.1 & 0 \\ 0 & 0.1 \end{pmatrix}$
  and
  $\begin{pmatrix} 0.5 & 0 \\ 0 & 0.5 \end{pmatrix}$.}
  \label{fig:city-isotropic}
\end{figure}

Neighborhood centers can be generated by sampling from a multivariate normal
distribution
\[
\mathbf{x} \sim \mathcal{N}(\mu,\Sigma),
\]
where $\Sigma$ is the covariance matrix, which governs the dispersion of
neighborhood centers around a mean location. In an unconstrained setting,
isotropic urban expansion would be characterized by a diagonal covariance
matrix with equal eigenvalues, leading to a uniform spread in all directions
(see Fig.~\ref{fig:city-isotropic}).

However, in the presence of geographical constraints, urban expansion is often
anisotropic, favoring certain directions over others.

\paragraph{Eigenstructure of the Covariance Matrix.}

The covariance matrix $\Sigma$ is assumed to be symmetric and real, ensuring it
is diagonalizable with an eigen-decomposition of the form
\[
\Sigma = Q \Lambda Q^\top,
\]
where $Q$ is an orthogonal matrix whose columns are the eigenvectors of
$\Sigma$, and $\Lambda$ is a diagonal matrix containing the corresponding
eigenvalues.
\begin{itemize}
  \item The eigenvectors of $\Sigma$ define the main directions of urban
  expansion. For example, if growth follows a river, one eigenvector aligns
  with its direction.
  \item The eigenvalues control dispersion along each direction: larger values
  indicate greater spread, while smaller ones reflect constrained growth.
\end{itemize}

\begin{figure}[!h]
  \centering
  \includegraphics[width=4cm]{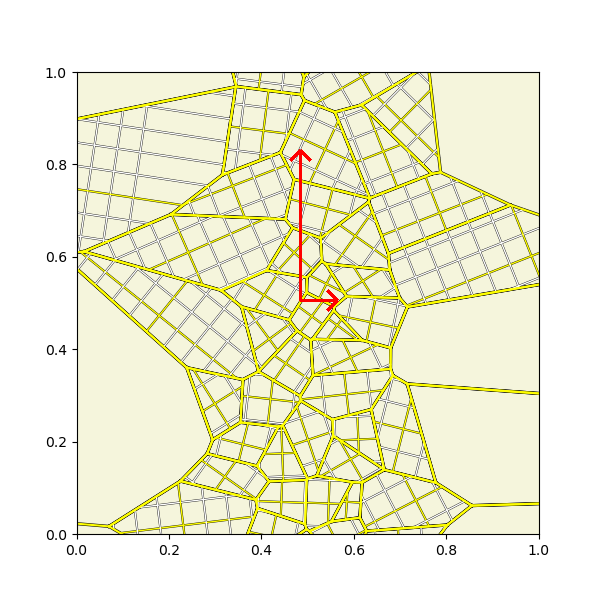}
  \caption{City generated with covariance matrix
  $\begin{pmatrix} 0.02 & 0 \\ 0 & 0.7 \end{pmatrix}$.}
  \label{fig:city-aniso}
\end{figure}

By adjusting the eigenstructure of $\Sigma$, the model can represent various
expansion patterns. In particular, higher eigenvalues can be assigned along
preferred directions such as rivers or coastlines (see Fig.~\ref{fig:city-aniso}).

\section{Discussion}

The hyperfractal city model introduced in this paper offers a powerful and
flexible framework for generating synthetic urban environments that incorporate
both geometric complexity and realistic traffic distributions. Nevertheless,
several limitations remain and suggest directions for future research.

First, the current model assumes uniform traffic density on each segment of a
given level, which captures average behavior but fails to reflect local
variability that occurs in real cities. Moreover, the districts in the model
are generated independently, whereas actual cities exhibit strong
interdependencies between neighborhoods due to shared infrastructure, mobility
flows, and historical development patterns. Another simplification is the use
of idealized Manhattan grids; while effective for modeling planned urban zones,
such grids may not represent more organic or irregular city layouts typically
found in older urban centers.

Despite these limitations, the model presents numerous opportunities for
practical applications, including the testing of routing algorithms and the
generation of synthetic datasets for machine learning. The model can also be
extended to a three-dimensional version, which can represent, for instance, ant
networks. It also opens up intriguing theoretical questions. For example, the
relationship between the fractal dimension of the measure and the efficiency of
traffic propagation or information dissemination on the network remains largely
unexplored. Lastly, the use of anisotropic Gaussian distributions to model
urban sprawl along natural constraints such as rivers or coastlines suggests
that the framework could be extended to incorporate temporal dynamics and
geographic data, making it even more applicable to real-world urban studies.


\bibliographystyle{splncs04}
\bibliography{bibliographie}

\end{document}